\documentclass[10pt, conference]{IEEEtran}

\usepackage{cite}
\usepackage[utf8]{inputenc}
\usepackage{amsmath,amssymb,amsfonts}
\usepackage{algorithmic}
\usepackage{graphicx}
\usepackage{textcomp}
\usepackage{xcolor}
\usepackage{siunitx}
\usepackage{dblfloatfix}
\usepackage{multirow}
\usepackage{array}
\usepackage{soul}
\usepackage{float}
\usepackage{subcaption}
\usepackage{hyperref}
\usepackage{textcomp}

\usepackage{pdfpages}

\begin{document}

\begin{titlepage}
\includepdf{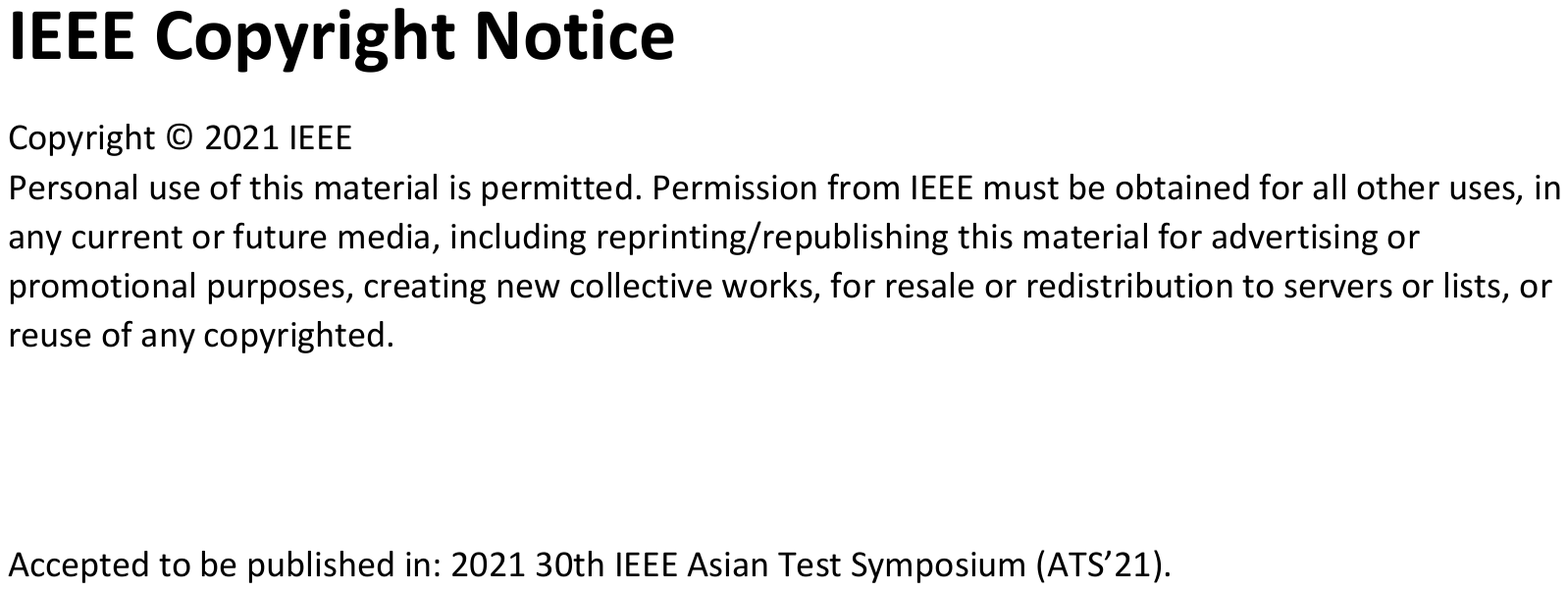}
\end{titlepage}
	
	\title{Side-Channel Attacks on Triple Modular Redundancy Schemes}
	
	\author{
		\IEEEauthorblockN{Felipe Almeida, Levent~Aksoy, Jaan~Raik, and Samuel Pagliarini}
		\IEEEauthorblockA{Department of Computer Systems \\ Tallinn University of Technology, Tallinn, Estonia\\
			Email: \{felipe.almeida, levent.aksoy, jaan.raik, samuel.pagliarini\}@taltech.ee}
	}
	
	\maketitle
	
	\begin{abstract}
		Triple Modular Redundancy (TMR) is a well-known fault tolerance technique for avoiding errors in the Integrated Circuits (ICs) and it has been used in a wide range of applications. The TMR technique employs three instances of circuits realizing concurrently the same functionality whose outputs are compared through a majority voter. On the other hand, Side-Channel Attacks (SCAs) are powerful techniques to extract secret information from ICs based on the data collected from security critical operations. Over the years, the interplay between security and reliability is poorly studied. In this paper, we explore the performance of SCAs on the well-known Advanced Encryption Standard (AES) and its different realizations using the TMR technique. In this work, three implementations of the AES design under the TMR scheme are used and an SCA, which can collect power dissipation data from the physical netlist through simulations, is developed. The experimental results show that the TMR technique can increase the computation time of SCAs and more importantly, the use of functionally equivalent, but physically and structurally different instances in the TMR scheme can make it impossible for SCAs to discover the secret key.
		
	\end{abstract}
	
	
	\IEEEpeerreviewmaketitle
	
	\section{Introduction}
	
	\IEEEPARstart{A}s the semiconductor industry pushes the limits of transistor technology in a never ending pursuit of miniaturization, radiation effects have become a serious concern not only for aerospace and military applications, but also for terrestrial applications. Among many radiation effects an Integrated Circuit (IC) may suffer from, Single-Event Transients (SETs) and Single-Event Upsets (SEUs) \cite{Baumann2005} are widely studied. The underlying principle is that a charged particle, upon striking the IC, may cause shifts in voltage levels at combinational or sequential elements, creating SETs or SEUs, respectively.
	
	Over the years, many efficient techniques have been used to mitigate radiation effects \cite{Kasap2020}, often making the use of some notion of spatial or temporal redundancy \cite{nicolaidis, nsrec2012, jeon2012, almeida2012, pagliarini2021}. Triple Modular Redundancy (TMR), one of the most commonly utilized solutions, is a technique that employs three instances of a module and adds a majority voter at their outputs. The scheme, therefore, protects against any single fault in any of the modules. The TMR technique can be deployed with different levels of granularity \cite{almeida2012, pagliarini2021}, with diversification \cite{meubrodersubadestanford}, and also with approximation \cite{apptmr}. It also presents partial protection against multiple faults caused by single-event-induced charge sharing \cite{almeida2012}.
	
	However, when a fault tolerant circuit is implemented using the TMR or a similar technique, its resiliency against security vulnerabilities tends to be overlooked. Recently, the field of Hardware Security has received a lot of attention and defense techniques against various adversaries have been implemented for a range of circuits. Yet, the interplay between security techniques and fault tolerance methods is still poorly understood.
	
	In this paper, our aim is to highlight this interaction by taking an Advanced Encryption Standard (AES) crypto core as a case study. The reliability technique we are concerned with is TMR in its many forms. In this work, we realize three possible AES designs under the TMR scheme. While the first one has the identical AES instances, the second one includes the same AES instances optimized by the synthesis tool, and the third one has functionally equivalent, but physically and structurally different AES instances obtained by the clock gating~\cite{Hai2003} and retiming~\cite{Leiserson1983} design techniques. The security attack we are concerned with is the power analysis based side-channel attack (SCA). We develop our SCA which extracts the simulated power dissipation data from a physical implementation of the AES design and guesses the secret key using a statistical procedure. To the best of our knowledge, for the first time, we perform SCAs on an AES design implemented under the TMR scheme. We show that the discovery of the secret key in the design under a TMR scheme needs a large simulation data, increasing the computation time of the attack when compared to the single AES design. We also point out that the use of functionally equivalent, but physically and structurally different instances in a design implemented using a TMR technique increases the resiliency to SCAs due to different power traces in each AES instance, making it impossible to discover the secret key while the other designs under the TMR scheme are vulnerable to SCAs.
	
	The rest of this paper is organized as follows: In Section~\ref{sec:background}, we present the background concepts related to SCAs on crypto cores. The implementation of an AES crypto core and its different realizations using the TMR technique are described in Section~\ref{sec:aes_design}. We introduce our SCA based on power analysis in Section~\ref{sec:sca}. Experimental results are given in Section~\ref{sec:results} and finally, the paper is concluded in Section~\ref{sec:conc}.
	
	\section{Background}
	\label{sec:background}
	
	In an SCA, an adversary collects, in a non-invasive way, leakage data that can be used to discover private information and/or to gain privileged access to a circuit \cite{standaert2010}. Power consumption, timing, electromagnetic emanations, and even sound are examples of side-channels that can and have been exploited. Based on the analysis of this residual information, it is possible to perform an attack that breaks security assumptions. In this paper, our focus is on SCAs that exploit power traces as a form of leakage. The power analysis based SCAs can be categorized in three groups: i)~Simple Power Analysis (SPA); ii)~Differential Power Analysis (DPA); iii)~Correlation Power Analysis (CPA). SPA is a simple graph analysis of the power trace consumption over time. DPA uses statistical analyses at different times to correlate power consumption measurements with functionality. CPA uses a Hamming weight power model method~\cite{Mangard2007} for a more powerful attack.
	
	Crypto cores have been the typical targets of SCAs. In principle, the mathematics behind the crypto function is sound and cannot be broken by formal crypto analysis. However, the physical realization of the crypto function gives adversaries powerful information.   
	
	\begin{figure}[!t]
		\centering
		\includegraphics[width=1.0\linewidth]{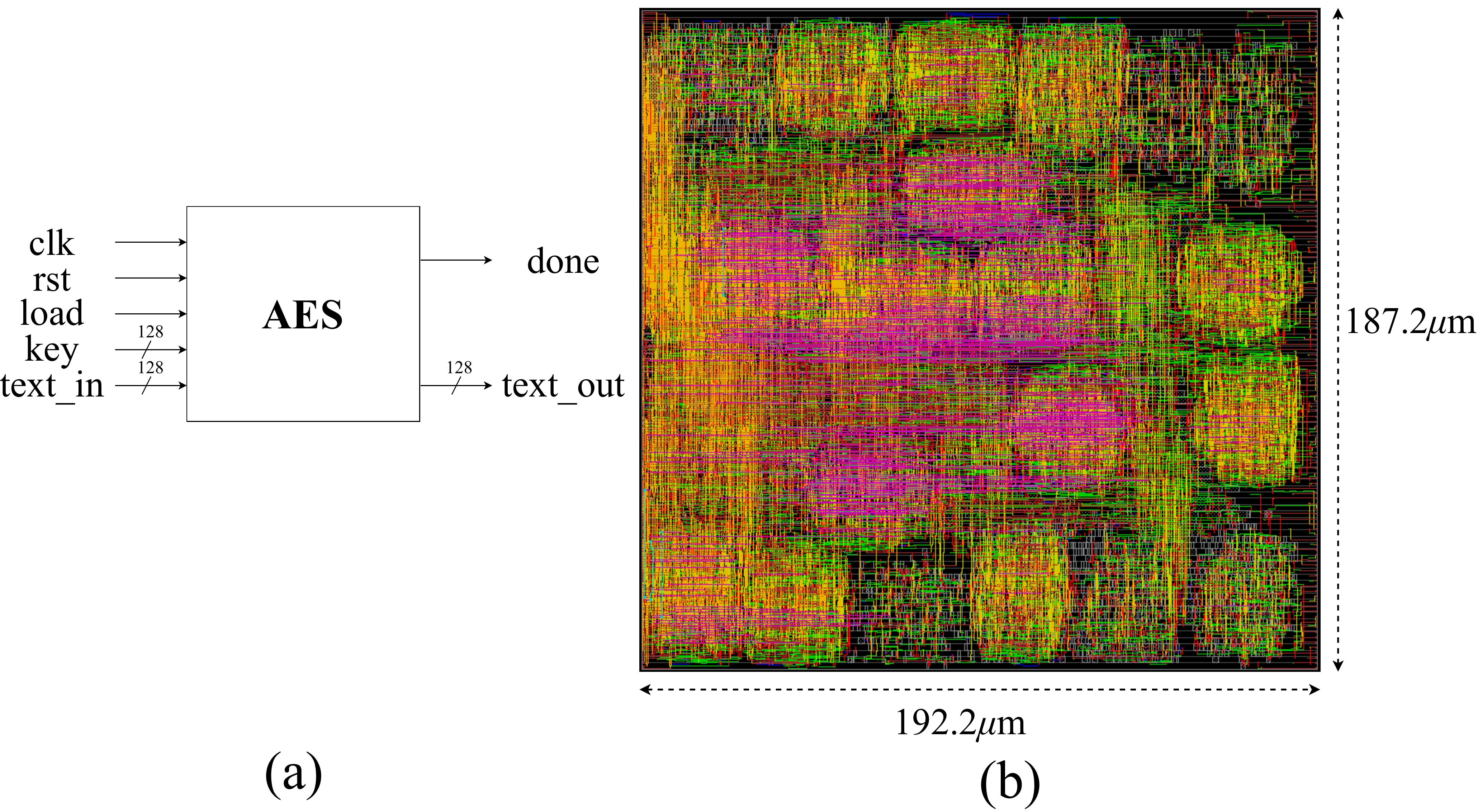}
		\caption{(a)~Block diagram of the AES circuit; (b)~its layout.}
		\label{fig:block_view}
		\vspace*{-4mm}
	\end{figure}
	
	In \cite{Maingot2009}, an evaluation of the sensitivity to DPA of several protected versions of an AES circuit is discussed. In \cite{Ors}, a power analysis attack on an AES hardware implementation is presented and an SCA is mounted on a physical device with the aid of a simple setup (scope and probes). The attack utilizes the power consumption during the first two clock cycles of the AES computation to discover the secret key. The reason for which the attack works is that in the considered AES implementation, an {\sc xor} operation between the plaintext and the secret key is executed in the first clock cycle. The result of this operation is saved in an \textit{intermediate register} in the second clock cycle. The adversary can devise a \textbf{hypothetical power model} to account for changes in the value of the intermediate register, i.e., the adversary can use bit changes in this register as a proxy for the behavior of the power consumption of the entire AES circuit. Even further, by simulation means, the adversary can analyse all possible changes the register might have, e.g., toggle count, in a cycle-accurate manner. This type of modeling is widely utilized in SCAs to discover the secret key in a device that implements AES. 
	
	Previous interactions between reliability and security can be found in mitigating hardware Trojans using the TMR technique. In~\cite{Mao2016}, an optimized graph partitioning of the TMR technique is used against hardware Trojan in a reconfigurable hardware and a fine-grain TMR architecture is presented to mitigate multiple faults and hardware Trojan insertion in~\cite{Gunti2017}. 
	
	This paper explores the performance of an SCA on the AES design under a TMR scheme. The proposed attack focuses on power consumption information leakage to discover the secret key in an AES crypto core. We assume that the AES core is meant for a high-dependability application and therefore, TMR has been applied to it. We also assume that the adversary has access to power traces of the circuit under attack. Furthermore, our approach emulates a physical attack by obtaining detailed power traces from physical synthesis. In practice, a real attack is more complicated because the environment, board, and package become sources of noise that have to be accounted for. We direct the readers to \cite{Ors} for more details on attack feasibility.
	
	\section{AES Crypto Core Implementation and its Realizations Using the TMR Technique}
	\label{sec:aes_design}
	
	As a case study for an SCA on a design under a TMR scheme, the AES crypto core is considered. Fig.~\ref{fig:block_view}(a) shows its block diagram. The AES circuit takes a 128-bit secret key ($key$) and a plaintext ($text\_in$) as inputs and produces a ciphertext as an output ($text\_out$). 
	
	\begin{figure}[!t]
		\centering
		\includegraphics[width=1.0\linewidth]{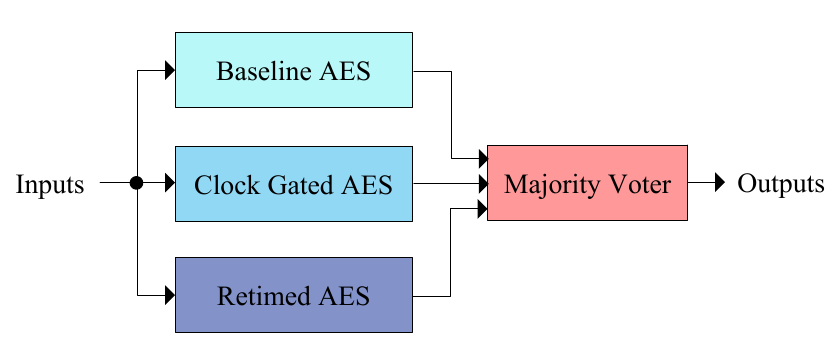}
		\caption{Structure of the {\sc aes\_tmr\_dif} design.}
		\label{fig:struct_dif}
		\vspace*{-6mm}
	\end{figure}
	
	\begin{figure*}[!hbt]
		\centering
		\includegraphics[width=1.0\linewidth]{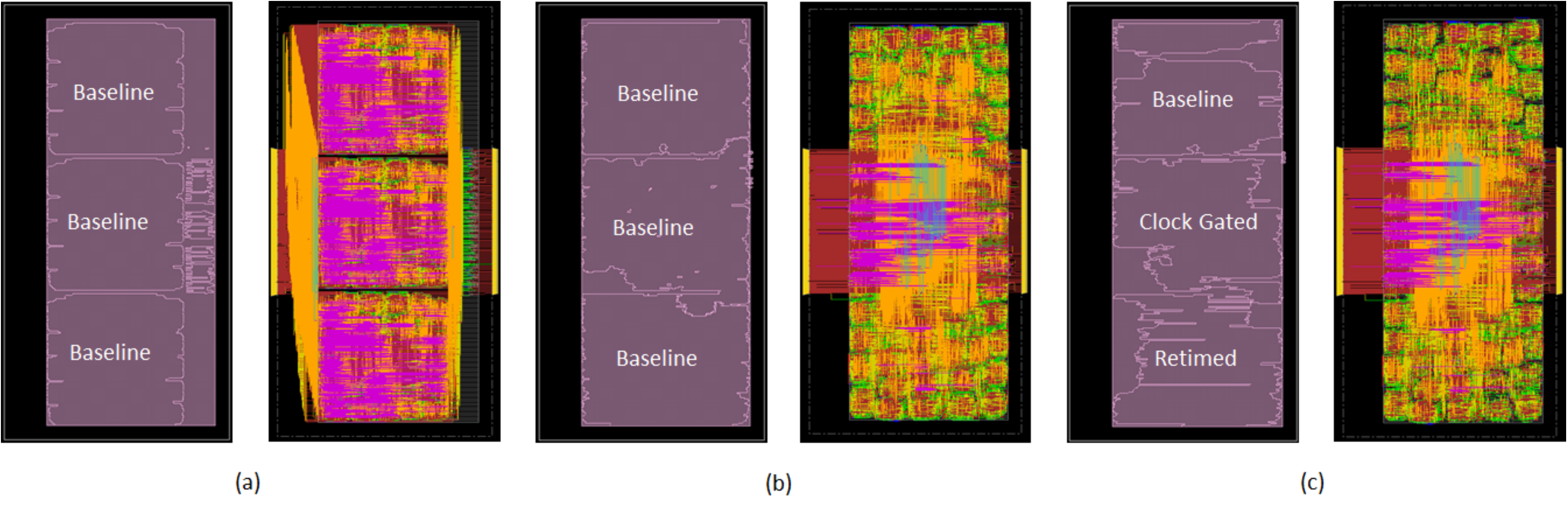}
		\caption{Amoeba views and layouts of TMR architectures:  (a)~{\sc aes\_tmr\_ide}; (b)~{\sc aes\_tmr\_opt}; (c)~{\sc aes\_tmr\_dif}.}\label{fig:TMR} 
		\vspace*{-4mm}
	\end{figure*}
	
	A 128-bit AES crypto core is obtained  from~\cite{aes_core} and implemented in a standard design flow. Initially, the logic synthesis of Verilog Hardware Description Language (HDL) codes of the AES circuit into a gate-level netlist is realized  using the Cadence Genus tool with a commercial 65 nm standard cell library when the target frequency is 500 MHz. Then, physical synthesis, including floorplanning, placement, clock tree, and routing, is performed by the Cadence Innovus tool. Fig.~\ref{fig:block_view}(b) presents the AES crypto core layout. This is our baseline implementation and is referred to as the \emph{single AES} in the rest of the paper. 
	
	The same AES crypto core was designed under a coarse-grain TMR architecture. Three different physical designs, called {\sc aes\_tmr\_ide}, {\sc aes\_tmr\_opt}, and {\sc aes\_tmr\_dif}, were considered. In the {\sc aes\_tmr\_ide} design, each instance in the TMR architecture is intentionally made identical: all cells and all metal routing lines are the same for all three instances. In the {\sc aes\_tmr\_opt} design, the physical synthesis tool is allowed to perform independent optimizations in these three instances if applicable. Finally, in the {\sc aes\_tmr\_dif} design, each instance is determined to be physically and structurally different, but functionally equivalent. To do so, we generated three AES crypto cores with different gate-level netlists. The first one is our baseline AES, the second one is obtained after applying the clock gating technique which is used to reduce power dissipation in parts of the circuit that are not being switched (and therefore, has an impact on SCA resiliency), and the third one is obtained after performing the retiming technique which moves the relative location of latches and registers, primarily to improve performance. In the {\sc aes\_tmr\_dif} design, the synthesis tool is also allowed to perform logic optimizations. The structure of the {\sc aes\_tmr\_dif} design is illustrated in Fig.~\ref{fig:struct_dif}.
	
	Figure~\ref{fig:TMR} presents the amoeba and physical layout views of the AES designs under the TMR scheme. Observe from Figure~\ref{fig:TMR} that the {\sc aes\_tmr\_ide} design includes three identical instances of the AES design, the {\sc aes\_tmr\_opt} design has three instances of the AES design structurally very close to each other, but with different number of cells and routes, and the {\sc aes\_tmr\_dif} design includes three physically and structurally different instances of the AES design. Note that all these TMR designs have the same timing constraints, core area, and pinouts for the sake of a fair comparison. 
	
	\section{Proposed Side-Channel Power Analysis Attack}
	\label{sec:sca}
	
\begin{figure}[!t]
		\centering
		\includegraphics[width=1.0\linewidth]{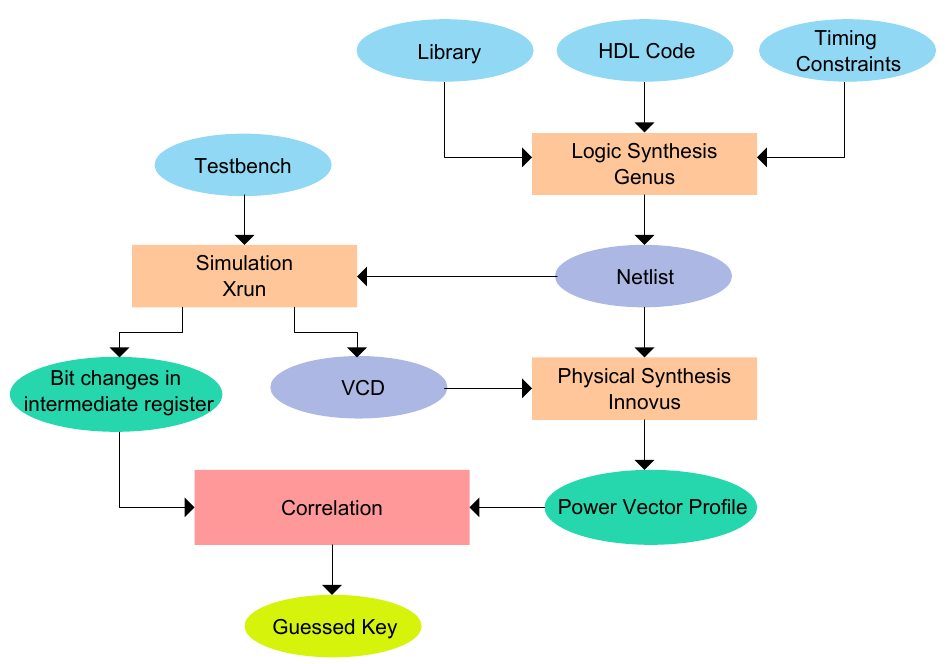}
		\caption{Flow of the side-channel power analysis attack.}
		\label{fig:sca_flow}
		\vspace*{-6mm}
	\end{figure}

	The flow of our side-channel power analysis attack is illustrated in Fig.~\ref{fig:sca_flow}. Compared to the traditional IC design flow, extra steps were included to enable our attack. To cope with the exponential size of all possible keys i.e., $2^{128}$, the simulation data is obtained for $L$-bits of the 128-bit secret key, where $L$ is set to 8 in our experiments. In the text and results that follow, without loss of generality, we perform attacks on 8 bits of the secret key at a time. The same attack can be repeated 16 times to uncover the entire 128-bit secret key. 
	
	In our attack, initially, logic synthesis is performed on the design using the timing constraints and design library by the Cadence Genus tool and the gate-level netlist is obtained. Then, this gate-level netlist is simulated using the Cadence Xrun tool under the given test-bench. Because $L$ is 8, this netlist is instantiated 256 times in the test-bench, i.e., one instance for each possible 8-bit key. One output of the simulation is the number of bit-changes in the intermediate register of the AES design, which stores the secret key in the first and second clock cycles as described in~\cite{Ors}, under all possible values of the $L$-bit key. Note that the number of bit-changes is a \mbox{high-level} representation of power dissipation. Another output of the simulation is the Value Change Dump (VCD) file which annotates any changes in any signals of the design along with the time of change. 
	
	Then, the gate-level netlist is passed through the physical synthesis performed by the Cadence Innovus tool. This tool reads the VCD file and generates a vector-based dynamic power report for any time window of interest under all possible values of the $L$-bit key. This power estimation is a good representation of the power dissipation of the fabricated chip because it takes into account parasitic information from extraction and representative input patterns from simulation\footnote{For readers with IC design background, we clarify that we utilize the Voltus power analysis engine of Innovus with VCD and Standard Delay Format (SDF) files. We ask the tool to generate a power estimation at every 1ns to oversample the 500 MHz frequency of operation of the circuit. This matches the capability of an adversary equipped with a typical oscilloscope.}. We obtain the \textit{power data set} which is computed as the difference of the power dissipation values of the AES crypto core in the first and second clock cycles as described in~\cite{Ors}. To obtain these simulation and power data sets, 1000 randomly generated plaintexts were used. 
	
	Finally, for each possible key, the Pearson Correlation Coefficient (PCC) is computed between the simulation and power data sets, and the one that leads to the maximum PCC value is determined to be the guessed key. 
	
	In order to make the Cadence tools work in harmony, the flow illustrated in Fig.~\ref{fig:sca_flow} is automated using Python scripting. Note that the runtime to discover the 8 bits of the secret key in the single AES design is approximately 2 hours for 1000 plaintext inputs. The majority of the runtime is spent during the generation of the power vector profile by the Cadence Innovus tool and the correlation calculation is much simpler in comparison.
	
	\section{Experimental Results}
	\label{sec:results}
	
	In this section, we first present the synthesis results of AES designs described in Section~\ref{sec:aes_design} and then, show the results of our attack introduced in Section~\ref{sec:sca} on these designs.
	
	\begin{table}[t]
		\centering
		\caption{Physical synthesis results of the single AES and its realizations using the TMR technique.} \label{tab:report_design}
		\begin{tabular}{|l||c|c|c|c|}
			\hline
			{\centering{\bf Design}} & {\centering{\bf gate}} & {\centering{\bf FF}} & {\centering{\bf area}} & {\centering{\bf power}} \\
			\hline \hline
			Single AES                & {11782}  & {530}  & {33.63}  & {9.44}  \\
			{\sc aes\_tmr\_ide}       & {35919}  & {1590} & {103.62} & {42.60} \\
			{\sc aes\_tmr\_opt}       & {35020}  & {1590} & {103.92} & {29.09} \\
			{\sc aes\_tmr\_dif}       & {30124}  & {1584} & {79.08}  & {50.51} \\
			\hline
		\end{tabular}
	\end{table}
	
	\begin{table}[t]
		\centering
		\caption{Physical synthesis results of each instance of the {\sc aes\_tmr\_dif} design.} \label{tab:diver}
		\begin{tabular}{|l||c|c|c|c|}
			\hline
			{\centering{\bf Instance}} &
			{\centering{\bf gate}} & {\centering{\bf FF}} & {\centering{\bf area}} & {\centering{\bf power}} \\
			\hline  \hline 
			Baseline AES    & {9826}  & {530} & {25.89} &  {15.84} \\
			Clock Gated AES & {9853}  & {530} & {25.91} &  {15.12} \\
			Retimed AES     & {10187} & {524} & {25.88} &  {16.73} \\
			\hline
		\end{tabular}
		\vspace*{-4mm}
	\end{table}
	
	\begin{figure*}[t]
		\centering
		\includegraphics[width=1.0\linewidth]{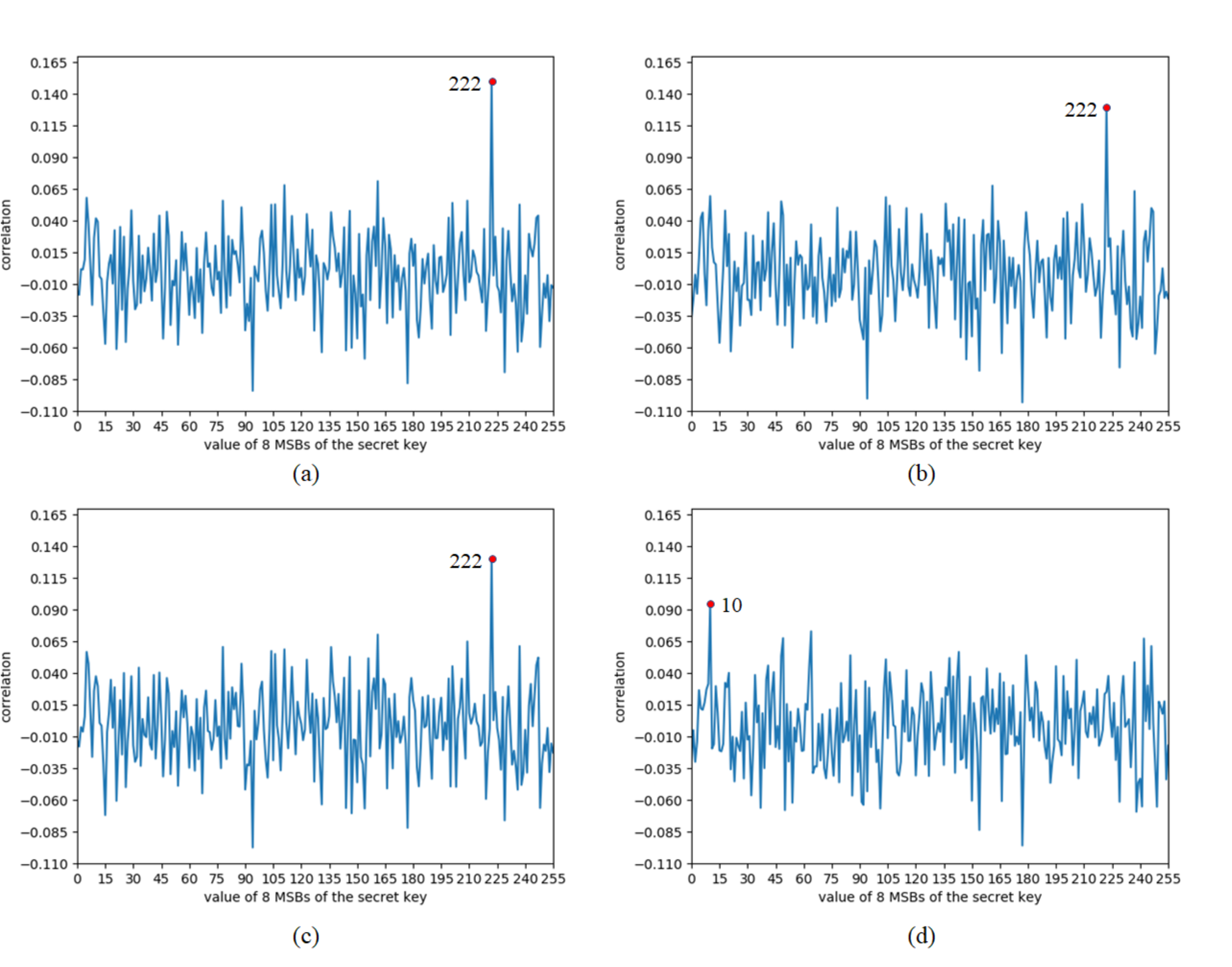}
		\vspace*{-4mm}
		\caption{Correlation between the simulation and power data sets when the 8 MSBs of the secret key was set to 222: (a)~single AES, (b)~{\sc aes\_tmr\_ide}, (c)~{\sc aes\_tmr\_opt}, and (d)~{\sc aes\_tmr\_dif}.}
		\vspace*{-4mm}
		\label{fig:correlation}
	\end{figure*}
	
	Table~\ref{tab:report_design} presents the physical synthesis results of the single AES design and its realizations using the TMR technique. In this table, \textit{gate} and \textit{FF} denote the number of gates and \mbox{flip-flops}, respectively and \textit{area} and \textit{power} stand for the total area in $\mu m^2$ and power dissipation in mW, respectively. Note that the \textit{area} includes all the cells and routes in the design and the clock frequency for all these designs is 500~MHz.
	
	Observe from Table~\ref{tab:report_design} that the realizations of the AES design using the TMR techique have around $3\times$ larger hardware complexity than the single AES design as expected. The {\sc aes\_tmr\_ide} and {\sc aes\_tmr\_opt} designs have hardware complexity very close to each other. On the other hand, the use of clock gating and retiming techniques in the {\sc aes\_tmr\_dif} design and logic optimizations allowed in the synthesis tool lead to around 23\% reduction in area with respect to other TMR realizations. However, power dissipation is increased $1.2\times$ and $1.7\times$ in this design with respect to the {\sc aes\_tmr\_ide} and {\sc aes\_tmr\_opt} designs, respectively. 
	
	Table \ref{tab:diver} shows the physical synthesis results of each AES instance of the {\sc aes\_tmr\_dif} design. Observe from Tables~\ref{tab:report_design} and~\ref{tab:diver} that all the AES instances have less complexity than the single AES design due to the logic optimizations performed by the synthesis tool. Observe from Table~\ref{tab:diver} that although these instances are physically and structurally different from each other, they have similar hardware complexity in terms of area.
	
	Our attack is run on the single AES and its realizations under the TMR scheme when 10 randomly generated 8 Most Significant Bits (MSBs) of the secret key are used. We note that in each experiment with a different secret key, our attack guessed the correct key in the single AES design and the {\sc aes\_tmr\_ide} and {\sc aes\_tmr\_opt} designs, but guessed the wrong key in the {\sc aes\_tmr\_dif} design. As an example, Fig.~\ref{fig:correlation} presents the PCC value for each possible key for the 8 MSBs of the secret key which was set to 222 under all AES designs. In these figures, the red dot denotes the key guessed by the attack which has the maximum correlation value. Observe from Fig.~\ref{fig:correlation} that the proposed SCA can discover the secret key in the single AES design and the {\sc aes\_tmr\_ide} and {\sc aes\_tmr\_opt} designs. The correlation value of the correct key in these designs are significantly larger than those of the wrong keys. However, our attack guesses a wrong key in the {\sc aes\_tmr\_dif} design, i.e., 10, and the correlation value of the guessed key is very close to those of other keys including the correct key. This experiment clearly indicates that the use of physically and structurally different AES designs under a TMR scheme increases the resiliency to  SCAs significantly, making the attack to guess a wrong key. This is simply because of different power traces in each AES instance under the {\sc aes\_tmr\_dif} design.
	
	\begin{figure*}[t]
		\centering
		\includegraphics[width=0.95\linewidth]{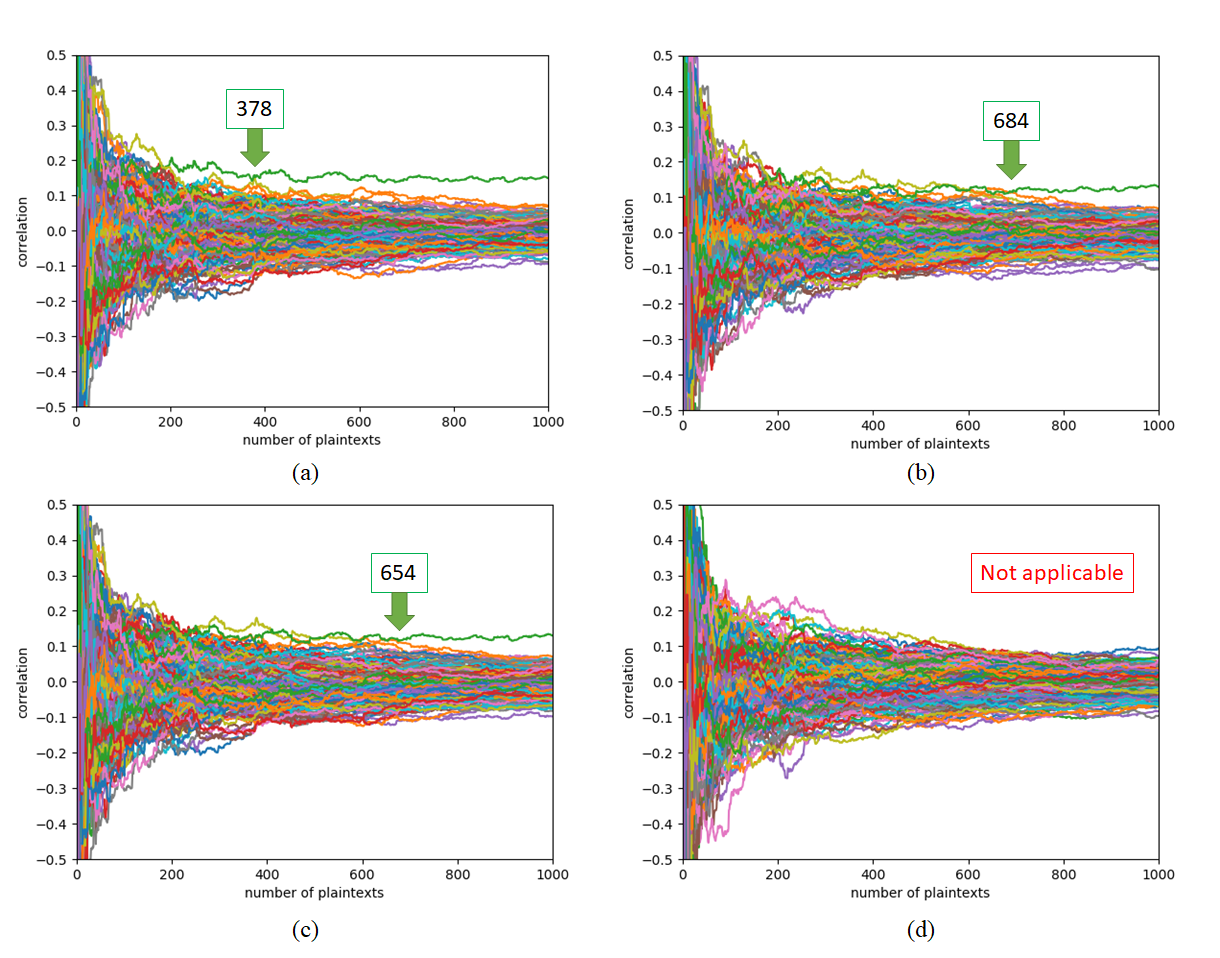}
		\vspace*{-2mm}
		\caption{Number of plaintexts necessary to discover the 8 MSBs of the secret key when they were set to 222: (a)~single AES, (b)~{\sc aes\_tmr\_ide}, (c)~{\sc aes\_tmr\_opt}, and (d)~{\sc aes\_tmr\_dif}.}
		\vspace*{-2mm}
		\label{fig:notexts}
	\end{figure*}
	
	Fig.~\ref{fig:notexts} presents the minimum number of plaintexts required to discover the 8 MSBs of the secret key. Observe from Fig.~\ref{fig:notexts} that the number of plaintexts required in the {\sc aes\_tmr\_ide} and {\sc aes\_tmr\_opt} designs is larger than the one required in the single AES design. This experiment clearly shows that the use of a TMR technique can increase the computational effort in SCAs. We note that our attack could not guess the value of 10 randomly generated secret keys of the {\sc aes\_tmr\_dif} design correctly even 2000 plaintexts were used. In this case, the run-time of our attack was almost doubled.
	
\begin{figure}[!t]
		\centering
		\vspace*{-32mm}
		\includegraphics[width=0.95\linewidth]{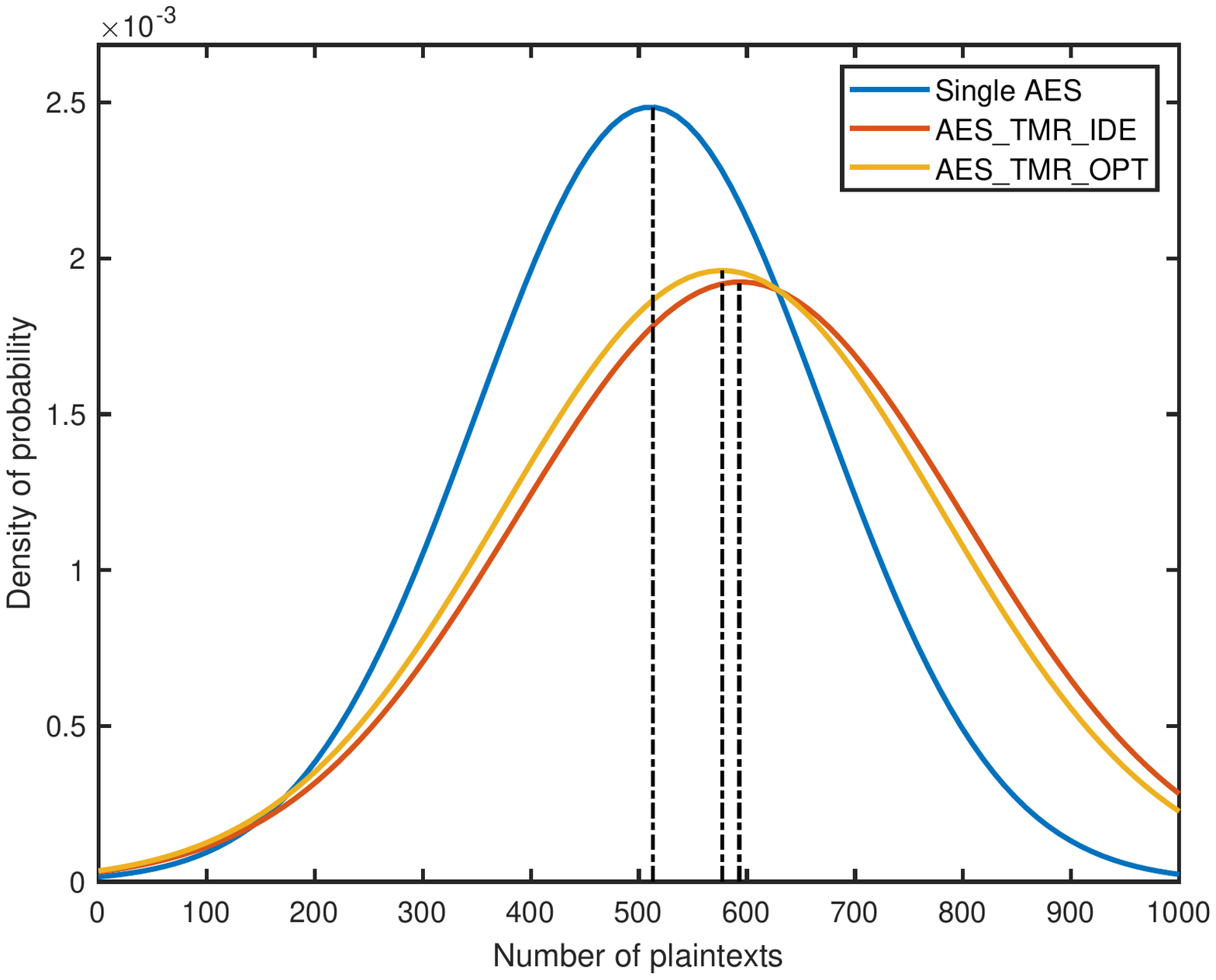}
		\vspace*{-28mm}
		\caption{Normal distribution on the number of plaintexts required to discover the secret key.}
		\vspace*{-6mm}
		\label{fig:normal}
	\end{figure}

	Finally, Fig.~\ref{fig:normal} presents the normal distribution on the minimum number of plaintexts required to discover the correct value of the 8 MSBs of the secret key obtained for successful attacks under all AES designs, except the {\sc aes\_tmr\_dif} design. Note that the dashed line points the average value of the number of plaintexts under the related AES design. Observe from Fig.~\ref{fig:normal} that while the {\sc aes\_tmr\_ide} and {\sc aes\_tmr\_opt} designs have a distribution very close to each other, their average values are larger than that of the single AES design. This experiment indicates that the use of a TMR technique can increase the number of plaintexts required to discover the secret key, increasing the computational effort in SCAs. 
	
	\section{Conclusion}
	\label{sec:conc}
	
	This paper demonstrated how a fault tolerance technique interferes with security, more precisely with the SCA resiliency, and showed how a TMR scheme with diversity can be leveraged to improve the resiliency of the design to SCAs. The experimental results pointed out that the use of a TMR technique can increase the number of plaintexts required to discover the secret key, increasing the computational effort in SCAs and the use of functionally equivalent, but physically and structurally different instances can make SCAs to guess a wrong key, increasing the resiliency of the design. As it stands, the use of reliability techniques to increase the security of a circuit is a largely unexplored territory. The possibilities for future avenues of research are plenty, including the study of redundancy schemes other than TMR and other crypto cores vulnerable to SCAs. 
	
	\section*{Acknowledgment}
	This work has been partially conducted in the project ``ICT programme'' which was supported by the European Union through the ESF. It was also partially supported by the Estonian Research Council grant MOBERC35.
	
	\bibliographystyle{IEEEtran}
	\bibliography{references}
	
\end{document}